\documentclass{aastex62}

\submitjournal{ApJ}
\usepackage{lipsum}
\usepackage{mwe}
\makeatletter
\let\saved@includegraphics\includegraphics
\AtBeginDocument{\let\includegraphics\saved@includegraphics}
\makeatother
\let\cite\citep
\usepackage{natbib}
\usepackage{adjustbox}

\shorttitle{Lunar Constraints on the Sun's Initial Rotation}
\shortauthors{Saxena et al.}

\begin{document}

\title{Was the Sun a Slow Rotator? - Sodium and Potassium Constraints from the Lunar Regolith}

\correspondingauthor{Prabal Saxena}
\email{prabal.saxena@nasa.gov}

\author{Prabal Saxena}
\affiliation{NASA Goddard Space Flight Center, Greenbelt, Maryland 20771, USA \\}
\affiliation{CREST II/University of Maryland, College Park, Maryland 20742, USA. \\}

\author{Rosemary M. Killen}
\affiliation{NASA Goddard Space Flight Center, Greenbelt, Maryland 20771, USA \\}

\author{Vladimir Airapetian}
\affiliation{NASA Goddard Space Flight Center, Greenbelt, Maryland 20771, USA \\}
\affiliation{American University, Washington, DC 20016, USA \\}

\author{Noah E. Petro}
\affiliation{NASA Goddard Space Flight Center, Greenbelt, Maryland 20771, USA \\}

\author{Natalie M. Curran}
\affiliation{NASA Goddard Space Flight Center, Greenbelt, Maryland 20771, USA \\}
\affiliation{USRA, Columbia, Maryland, 21046, USA \\}

\author{Avi M. Mandell}
\affiliation{NASA Goddard Space Flight Center, Greenbelt, Maryland 20771, USA \\}



\begin{abstract}

While the Earth and Moon are generally similar in composition, a notable difference between the two is the apparent depletion in moderately volatile elements in lunar samples.  This is often attributed to the formation process of the Moon and demonstrates the importance of these elements as evolutionary tracers.  Here we show that paleo space weather may have driven the loss of a significant portion of moderate volatiles, such as sodium and potassium from the surface of the Moon.  The remaining sodium and potassium in the regolith is dependent on the primordial rotation state of the Sun.  Notably, given the joint constraints shown in the observed degree of depletion of sodium and potassium in lunar samples and the evolution of activity of solar analogues over time, the Sun is highly likely to have been a slow rotator. Since the young Sun's activity was important in affecting the evolution of planetary surfaces, atmospheres, and habitability in the early Solar System, this is an important constraint on the solar activity environment at that time.  Finally, since solar activity was strongest in the first billion years of the Solar System, when the Moon was most heavily bombarded by impactors, evolution of the Sun's activity may also be recorded in lunar crust and would be an important well-preserved and relatively accessible record of past Solar System processes.

\end{abstract}


\section{Introduction} \label{sec:intro}

The evolution of the Sun's magnetic activity throughout the history of the Solar System is a key factor in understanding the past and current state of surfaces and atmospheres of planets in the inner Solar System. Solar activity could have played an important role in the habitability of a number of planets\cite{2018NatAs...2..448A, 2018arXiv180704776D}, including Earth\cite{2016NatGe...9..452A}, and may also have affected the evolution of planetary atmospheres and surfaces by influencing atmospheric loss\cite{Lammer2018} and chemistry\cite{2013GeoRL..40.1237T}. Indeed, evidence from meteorites suggests a period of higher solar activity early in the Solar Systems' history\cite{1987ApJ...313L..31C, 2005Natur.437..385A, 2018NatAs...2..709K}.  

Data from the Kepler space telescope \cite{2010Sci...327..977B} on the activity of solar analogues has provided important evidence on different pathways the Sun may have followed with respect to stellar activity.  Studies suggest that flare activity is greater for stars with shorter rotational periods, and that at these higher rotational velocities, Sun-like stars produce a significantly greater frequency of large, higher energy flares than Sun-like stars with rotational periods similar to our present day Sun \cite{2013ApJS..209....5S, 2013ApJ...771..127N}.  The Sun's rotation period is expected to have evolved after the zero age main sequence, with a shorter rotation period early in its lifetime that gradually slowed due to loss of angular momentum. The initial rotational velocity of the Sun is unknown and is critical as it would control the amount of magnetic flux emitted by the star, and as a result, flare, and consequently coronal mass ejection activity from the Sun early in the Solar Systems' history.  Semi-empirical models of evolution of the rotational behavior of solar-type stars have been successful at matching distributions of rotational periods observed for star forming regions and young open clusters corresponding to a wide range of ages \cite{2013A&A...556A..36G, 2016A&A...587A.105A}.  A useful representation of the evolutionary pathways from these studies tracks changes in rotation of three types of rotational classes for Sun-like stars: slow, medium and fast rotators.  These classes correspond to the 25th, 50th, and 90th percentiles of the rotation rates observed in the statistical sample in each of those clusters\cite{2013A&A...556A..36G}.

\section{Constructing an Earth-Moon System Incident CME Frequency History} \label{sec:CMEFreq}

\begin{figure}[t!]
\begin{center}
\includegraphics[scale=0.5]{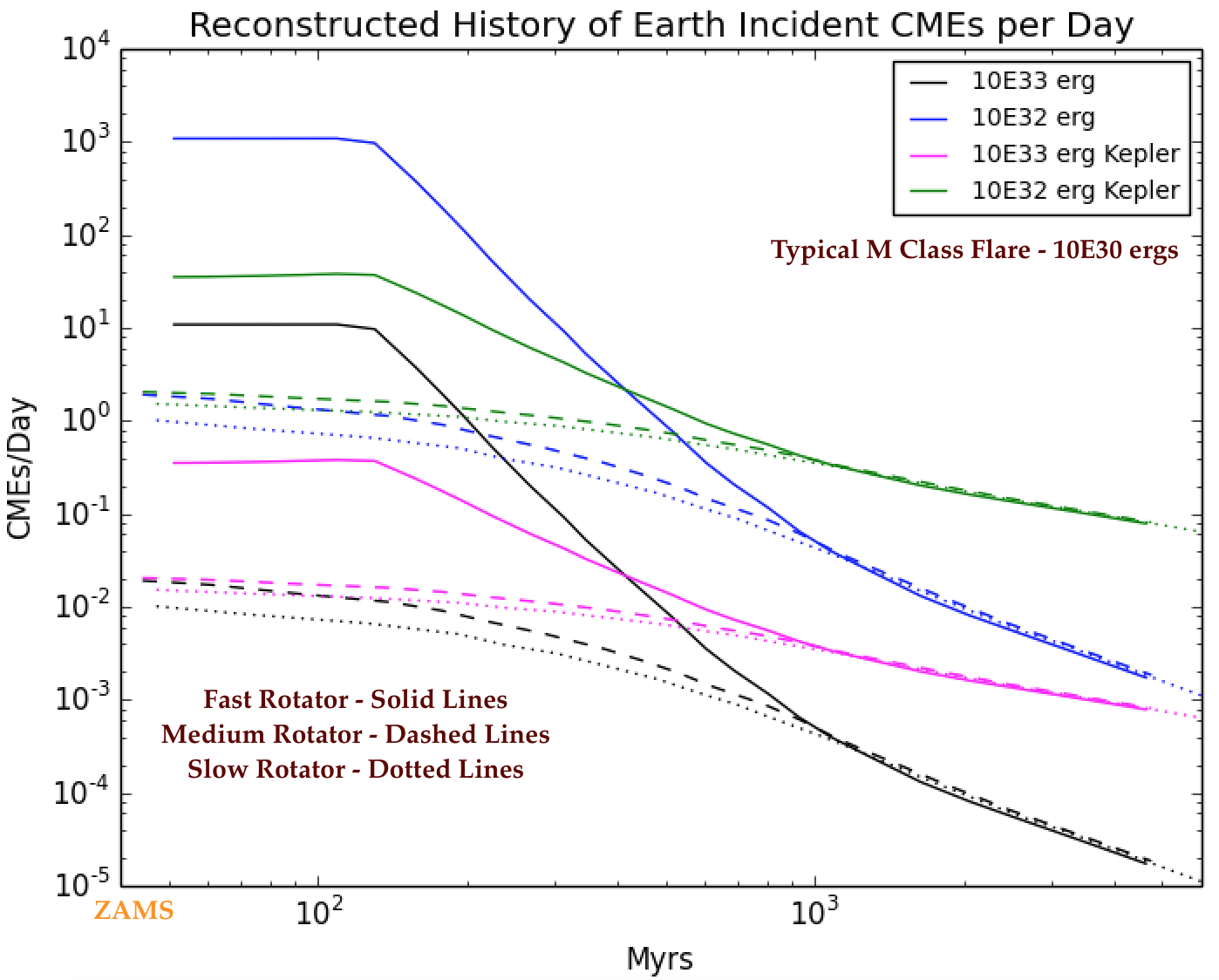}
\caption{\textit{A Reconstructed Flare/Coronal Mass Ejection history of the Sun for different rotation scenarios. It combines solar analogue flare rates from Kepler telescope\cite{2013ApJ...771..127N} observations with semi-emprical models of rotation rate vs age for Sun-like stars\cite{2013A&A...556A..36G}.  Flare energies correspond to X10 and X100 flares and frequencies are given for a strictly Kepler based rotation-flare rate relationship and for one calibrated by Earth geological records.}}
\label{fig:cme_frequency_time}
\end{center}
\end{figure}

While the initial rotational state of the Sun is unknown, the importance of constraining this history becomes apparent when contextualizing it with these different potential pathways. Using the rotation-flare relation\cite{2013ApJ...771..127N} for Sun-like stars identified in Kepler data and the association of flares with CMEs, we have reconstructed Earth-Moon system incident coronal mass ejection frequency histories for the Sun for the three different classes of rotational evolution in figure \ref{fig:cme_frequency_time}.  Figure \ref{fig:cme_frequency_time} shows how the different rotator cases for the Sun would have resulted in a different frequency of high energy Earth-Moon system incident CMEs (using flare frequencies and flare-CME relations described in this section). 

\begin{table}[]
\centering
\caption{Interplanetary Coronal Mass Ejection Properties}
\begin{tabular}{|c|c|c|c|c|c|}
\hline
 ICME - Flare Strength  & Velocity (km/s) & particle density (cm$^{-3}$) & f(He++) & Alpha/proton & O/He  \\
\hline
ICME - C Class                      & 400             & 10                     & 2\%     & 0.02         & 0.025 \\
\hline
ICME - M Class                      & 1000            & 30                     & 2\%     & 0.02         & 0.025 \\
\hline
ICME - X Class                      & 3000            & 100                    & 2\%     & 0.02         & 0.025 \\
\hline
ICME - Superflare \textgreater{}X10 & 6000            & 1000                   & 2\%     & 0.02         & 0.025 \\
\hline
\end{tabular}%
\label{table:solarprop}
\end{table}

This rate was derived by combining a rotation rate to flare frequency relation for Sun-like stars observed by the Kepler Space Telescope with empirically validated stellar rotation versus age relations for different classes of rotators for Sun-like stars.  The rotation-flare frequency relation is taken from the flare frequency versus brightness variation period bin bar graph in figure 7 from a study by Notsu et al.\cite{2013ApJ...771..127N} - the data uses Kepler data of Quarters 1-6.

We fit the data using a power law for rotation periods both less than and larger than 0.99 days (which is where the fit changes from flat to exponential).  The fit for periods larger than 0.991 days is the more significant one as it represents the relevant fit for vast majority of nearly all the rotator cases examined here (with an exception for a very short period of time before ZAMS for the fast rotator, which is not relevant to the study and is not included in the reconstructed image).  The fit for periods less than 0.991 days is also noted as being incomplete due to saturation at periods less than a few days.  As a result, the fits for this shorter period (y = 0.183x + 0.732 for the strictly Kepler relation and y = 0.183x + 25.2 for the relation calibrated by Earth geochemical records) are fairly flat with an intercept required to preserve continuity at 0.991 days.  In these fits, y represents the frequency of superflares as defined by Notsu et al.\cite{2013ApJ...771..127N} while x represents the brightness variation period taken to represent rotation rate of the star.

The relations for the longer than 0.991 day periods are given for two different cases: 1) a strictly Kepler data given relation for the data in figure 7 (y = 0.898x$^{-1.92}$) and 2) a relation using the Kepler data but using a more conservative super$-$CME frequency\cite{2017arXiv170903165G} with a lower occurrence rate for longer stellar rotation periods based on the Earth's record (y=24.5x$^{-4.12}$).  Of note is that we use the conservative estimate for the calculations used for depletion plotted in the paper.  While this means peak CME frequency rates are higher in the second case, the overall integrated number of CMEs over the age of the Sun is lower than in the first case since the strictly Kepler relation has a significantly higher flare frequency after $\sim$3.5Gyr.  To plot the CME frequency histories for the different rotator cases, we then combine the above relations with the rotation versus time evolution graph given in figure 3 of Gallet and Bouvier\cite{2013A&A...556A..36G}, which plots the rotation evolution of the empirically validated reconstructed rotator cases.  However, this only gives us a solar flare frequency (and in particular a superflare frequency for flares of an energy described in the Kepler study\cite{2013ApJ...771..127N}) to age relation for the different rotator cases.  Extracting an Earth-Moon system incident CME frequency history is predicated upon a number of assumptions that are described in the following paragraph.

Determining the flare to CME association rate is one of the most important assumptions we have to make given the sparse data available on stellar CME occurrence rates.  Observational evidence of CMEs, particularly on Sun-like stars, is difficult to obtain since such observations are inherently fortuitous given the projected doppler shifted spectroscopy that is used as a proxy\cite{2001ApJ...560..919B, 2017IAUS..328..198K}. We assume that nearly every one of these high energy flares produces a CME, based upon the 100\% solar CME to flare association rate for the X-ray peak flux values that correspond to flares of the plotted energies\cite{2009IAUS..257..233Y}.  Indeed, from figure 7 of \citet{2009IAUS..257..233Y}, even M class flares (which are more than an order of magnitude less energetic than the weakest flares we consider) have a CME association rate of $\sim80\%$ based on peak X-ray flux.  Even if we assumed this lower association rate and linearly scaled CME frequencies by it, it would not change the conclusions of this study. We conservatively assume an isotropic geometric distribution of CMEs when determining how many are incident on the Earth-Moon system, despite the likelihood that CMEs are likely to be preferentially confined to the solar equatorial plane.  We also use the same geometric model for CME morphology as given in \citet{2016NatGe...9..452A}, including angular cone width.  Despite the differing energies of the flares and CMEs that are of interest, empirical evidence \cite{2004JGRA..109.7105Y} suggests a 90 degree cone width assumption is reasonable and even a more conservative 60 degree cone width would only attenuate frequencies by of 1.5 and wouldn't affect our study's broader conclusions. We conservatively only include CMEs whose flare energy is above the $10^{32}$ erg threshold. This ensures that we only capture those CMEs that are definitively above energy thresholds needed to avoid being confined by the putatively stronger dipole magnetic fields in the early Sun\cite{2017MNRAS.472..876O, 2018ApJ...862...93A}. Finally, the interplanetary medium into which any early CMEs may have been injected may also have been different, but given the uncertain nature of how such a medium may have differed, we neglect possible effects in this study. 

From figure \ref{fig:cme_frequency_time}, it is apparent that different initial solar rotation rates could have meant significantly different early space weather environments.  While frequency of CMEs incident on the Earth-Moon system converges to about the same value for each solar rotation case after $\sim$1 Gyr, prior to this time, variations can be very large. Regardless of whether the reconstruction is calibrated to the Kepler data alone or to a combination of Kepler data and geological records from the Earth (which is more conservative), a medium rotator Sun has a frequency of CMEs 10s of \% greater (with a maximum of $\sim80\%$) than the slow rotator case prior to 1 Gyr. For fast rotators, the difference is even more pronounced, as the frequency is at times orders of magnitude greater than either of the other solar rotation scenarios.  In all cases, for a period of almost 0.5 Gyr, the Earth-Moon system would have experienced a frequency of at least $\sim1$ CME passage per 2 days, with the corresponding CME energies approximately in the range of some of the most energetic CMEs ever recorded.

\section{Model Description and Space Weather Assumptions} \label{sec:ModSpaceWeath}

We then try to understand whether there is any signature of this space weather that may be retained by the staid surface environment of the early Moon.  With the exception of relatively short periods of time\cite{2017E&PSL.474..198S, NEEDHAM2017175}, the more frequent CMEs produced by the young Sun would have been directly incident on the lunar surface.  Studies of the Moon during the passage of a CME have shown the total material lost from the surface of the Moon increases greatly during such an event \cite{doi:10.1029/2012JE004070, doi:10.1029/2011JE004011}.  Surface ion flux increases during the CME, driving sputtering loss from the surface into the lunar exosphere and then largely to escape from the Moon.  

We examined whether elemental abundances from lunar samples could constrain the past activity of the Sun and conversely, how much past activity may have influenced the abundances of different elements.  We focused on two moderately volatile elements - sodium and potassium, which are observed to be depleted in Apollo samples and in lunar meteorites relative to terrestrial values \cite{2013ApJ...767L..12V, 1980LPSC...11..333T}. While these elements both appear to be depleted relative to terrestrial values and are not dominant constituents of lunar samples, they are abundant enough (ranging from $10^{-4}$ to $\sim$ $10^{-2}$ by weight) to be measured across different rock types and conveniently also represent exosphere components that are observable in the visible spectrum (enabling observational constraints on abundance/loss). 

To understand how passage of energetic CMEs drives loss from the lunar surface, we use an updated version of a previously created surface bounded exosphere generation model\cite{doi:10.1029/2011JE004011}.   Sputtering rates are calculated for different cases, including low and medium energy Interplanetary CMEs (ICMEs), high energy ICMEs (associated with a X class flare), and superflares (associated with a X10).  We use the last two for the loss from CMEs associated with a $10^{32}$ and $10^{33}$ erg flare.  The properties for the incident CMEs that are then input into the exosphere generation model are given in table \ref{table:solarprop}.  These values are obtained from empirical studies\cite{Gopalswamy2012} of recent CMEs and from extrapolation from those values in the most extreme case.

The full methods for the exosphere generation model and sputtering processes in the model are given in the previous paper\cite{doi:10.1029/2011JE004011} which studied passage of a CME on the exosphere. The model consists of a monte carlo model which tracks the migration of particles under the influence gravity and radiation pressure once it is released from the surface into the exosphere.  The particles are then tracked as they either ballistically escape to space, are photoionized or photodissasociated, or return to the surface as they either stick or fall into a cold trap.  The model records particle positions and velocities at user-defined times, flux to specific points on the surface, and loss rates.  Input processes and functions are described in  \citet{doi:10.1029/2011JE004011}, but we choose specific space weather particle data for this project that is relevant to the process studied here.  In particular, the composition of the solar wind is taken from data from the Ulysses\cite{doi:10.1029/1999JA000358} mission for the fast wind, slow wind, shock and magnetic bubble gas, and from a paper looking at ICME parameters\cite{2008ApJ...682.1289R} for C-, M- and X- class flares.  Ulysses' perihelion distance was greater than 1 AU, so it is unlikely that it provides an overestimate of CME parameters from it's in situ data.  Particle densities used in this study also match those cited in \citet{doi:10.1029/2012JE004070} for passage of a moderate energy CME at the Earth-Moon system.  Of note is that we use ion abundances for the solar wind, another choice that is likely to produce a conservative sputtering and loss rate.  In our results we also show values that only incorporate the lower energy CMEs we consider as the higher energy ones require extrapolation given the relatively low frequency of such events at present.  

The primary driver of the increased sputtering yields are the enhanced particle densities and velocities during a CME event (if we used the higher ion fractions associated with CMEs, this too would be an important factor).  We specifically focus on physical sputtering given that yields from this process are far greater than photon stimulated desorption and micrometeorite impacts for the relevant events.  Sputter yields of neutral elements are scaled to KREEP soils on the Moon and are taken from a paper\cite{2011NIMPB.269.1310B} that examined the effect of sputtering on the lunar regolith and exosphere.  We only include physical sputtering processes in our model due to the uncertainty in inputs for potential sputtering (this results in an underestimate of loss as potential sputtering often increases yields by a factor of 2).  Full sputtering values for Na, K and other elements are given in the same paper\cite{2011NIMPB.269.1310B}, as are the input ion fractions (the paper gives kinetic-sputtering yields of a KREEP surface by solar-wind protons and heavy ions as taken from \citet{Ziegler1985}).  We also neglect photon stimulated desorption (PSD) effects given that they are negligible relative to sputtering during a CME event - from table 4 in \citet{doi:10.1029/2011JE004011} it is evident that PSD is more than two orders of magnitude less efficient in ejecting sodium and potassium from the surface than sputtering (this is also evident from table \ref{table:sputterval} where multiplying values by $\sim$ 4E17 for the total surface area of the Moon in $cm^{2}$ also yields similar much larger values for sputtering).  As discussed later, given the relatively low fraction of regolith material that is meteoritic, and given the relatively minor contribution observed in the simulations in \citet{doi:10.1029/2011JE004011}, we also do not include micrometeorite effects as they are negligible versus sputtering escape.  Of note is that both of these processes would contribute more loss of sodium and potassium, strengthening the underlying conclusions or this study.

We then apply escape fractions for the individual elements based on their likelihood to stick, be ionized or escape as defined in the model\cite{doi:10.1029/2011JE004011}.  We use this to estimate the total amount of an element lost from the exopshere during passage of a CME.  We use a 95\% loss rate for Na\cite{doi:10.1029/2011JE004011} and two bounding cases of 50\% and 90\% loss for K since there is no specific value available (given that the lower bound escape fraction is closer to much heavier elements, that should be viewed as a conservative escape fraction\cite{KILLEN20182364}).  We verify that the total losses for an M-flare associated event matches those given in previous work\cite{doi:10.1029/2011JE004011}.  Sputtering yields for different elements before the escape fraction is applied are given in table \ref{table:sputterval}.

We can integrate element loss for each individual CME passage over time using frequencies given for different cases in figure \ref{fig:cme_frequency_time}.  Estimates of the total loss are then produced for different solar rotation cases. In figures \ref{fig:cme_frequency_timeNa}-\ref{fig:cme_frequency_timeK90} we show loss as a function of three different time and event scenarios.  All loss estimates use conservative frequencies from figure \ref{fig:cme_frequency_time}.  Red circles indicate total lost from the surface of the Moon due to the passage of $10^{32}$ erg flare associated CMEs over the Moon's history (conservatively beginning at 110 Myr, after a late forming impact\cite{2004Icar..168..433C} and crust solidification).  Yellow circles indicate total lost from the surface of the Moon due to the passage of both classes of CMEs. Green values are loss due to the passage of $10^{32}$ erg flare associated CMEs in the last $\sim$3.5 billion years, after the emplacement of the greatest volume of mare basalts\cite{doi:10.1130/SPE477}. Top circles represent 2 day CME passage duration while the bottom represent 1 day passage.   

Total loss estimates of an element due to incident CMEs in separate cases where CMEs associated with $10^{32}$ and $10^{33}$ erg flares are considered and also where only the less energetic CMEs are considered were made in order to assess the effect of stronger CMEs.  Since sputtering scales with mass, energy, and charge state of the incident ions, which are a function of CME energy, this helps extricate the effects of what are likely to be the more commonplace CMEs from those that are rarer and less well understood (indeed, this is why we neglect to include even more energetic CMEs).  In general, the more energetic CMEs contribute about 20\% of the total loss relative to the less energetic CMEs.  Thus, there is no major difference in total depletion with respect our study's overarching conclusions when considering only the less energetic CMEs, versus a combination of both CMEs.  Additional yield per event during more energetic CMEs is swamped by their significantly lower frequency.

We ignore loss and implantation due to the solar wind and co-rotating interacting regions - this conservative approximation is based on initial estimates which suggest loss due to the solar wind is about an order of magnitude smaller than the loss due to CMEs when integrated over time.  Sputtering rates of Na and K are more than $>$100 times greater for the lower energy CMEs tested versus rates during the ambient slow wind (see table \ref{table:sputterval} versus table 8 of \citet{doi:10.1029/2011JE004011}).  Given the higher frequencies of CMEs expected in the first 500 million years, loss from a similar wind would be on the order of 1-10\% relative to total loss (with the latter number being due to a stronger early solar wind).  We neglect loss due to co-rotating interacting regions due to uncertainty regarding their incidence frequency and parameters.  It is important to note that numerous assumptions have been made or processes neglected in order to remain conservative about the total elemental loss that may  have occurred - these may compound in a way that results in a somewhat significant underestimate of the total elemental depletion.  However, we prefer to be cautious on these assumptions as relaxing them would only further support the conclusions of the study.  The total loss due to CME passage integrated over time is given in table \ref{table:lossest}.  The table is the source of the depletion figures in the main body and breaks down losses by CME energy, element choice, escape fraction and time period.  Total loss over time is given with respect to the total integrated time after lunar crust formation (in a late Moon formation scenario with an assumed short time of $\sim$1000 yrs to flotation crust formation\cite{2017E&PSL.474..198S}) and also for total time starting from 1 Gyr after crust formation.  This additional time period was chosen since it is approximately when all the rotator cases converge to lower and similar incident CME frequencies and is also conveniently around the same time as the period of emplacement of the largest volume of mare basalts.

\section{Regolith and Surface Composition Assumptions and Constraints} \label{sec:SurfComp}

Total loss relative to the elemental reservoir of the regolith is also dependent on crustal assumptions.   The portion of the regolith from which sputtering drives loss is approximately the top 100 nm\cite{2007Icar..191..486W} to $\sim$cm for CMEs\cite{Mckay_7the}.  Regolith assumptions are critical to the study since they determine the reservoirs from which the moderate volatiles are lost. Additionally, creation and gardening of the regolith determines what portions of the regolith were exposed to the relevant space weather processes. Estimates of the depth of the loose, fragmented rock that compose this portion of the regolith generally range in the 4-5 meter range for mare regions and 10-15 meter depths for highland regions\cite{Mckay_7the}.
This is the portion of the surface that was most susceptible to loss due to solar activity because it was immediately created and churned by impacts to significant depth in the Moon's early history (indeed, even conservative models suggest this portion was largely created and churned in the 1st Gyr after formation\cite{1974LPSC....5.2365G}). The megaregolith bedrock below this was less likely to be exposed given this protective overlying layer and while fractured by larger impacts, remained relatively intact.  We assumed a regolith thickness of 5 meters for the 17\% of the surface corresponding to mare surfaces and 15 meters for the 83\% corresponding to highlands\cite{1975LPICo.234...66H}.  The choice of the upper values in those ranges results in a conservative estimate of depletion given the larger mass reservoir.  

\begin{figure}[t]
\begin{center}
\includegraphics[scale=0.5]{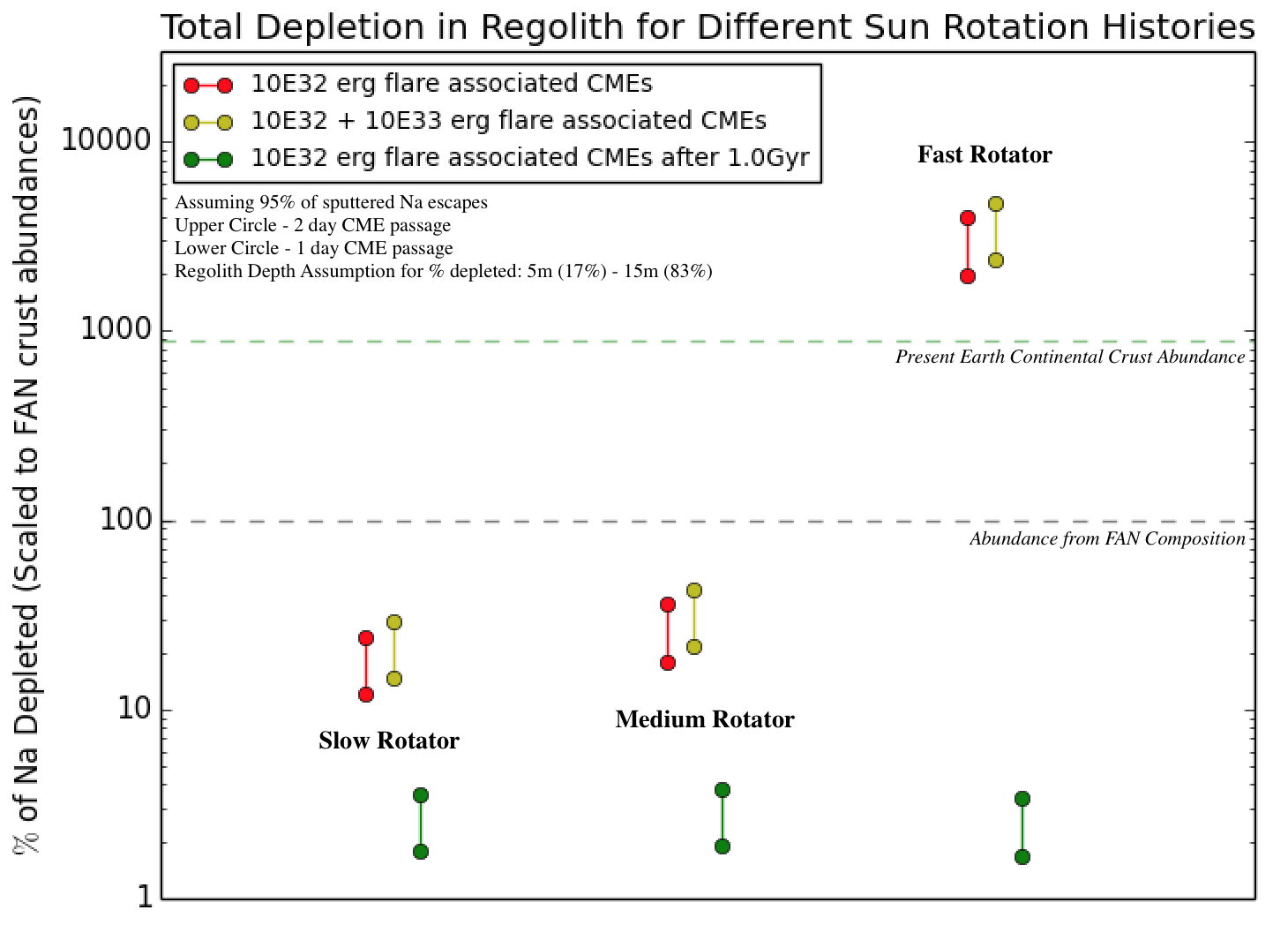}
\caption{\textit{Depletion of sodium from the lunar regolith for different solar rotation cases.  Different color values denote what CMEs were considered and are for the entire history post-crust formation unless stated otherwise.  Crust assumptions are given under the text box and crust composition lines are given by the different colored dashed lines.}}
\label{fig:cme_frequency_timeNa}
\end{center}
\end{figure}

We also choose a conservative value with respect to depletion for the average density of the regolith of 2400 kg/m$^{3}$.  The chosen regolith density value more accurately corresponds to modeled values for the surface density of the megaregolith\cite{doi:10.1002/2014GL059378}.  While the top 10's of cm of the regolith is significantly less dense (1500-2000 kg/m$^{3}$), we choose this value to obtain conservative estimates of depletion given what is known to be increasing density with depth\cite{doi:10.1029/2005RG000184}.  We then compare different solar rotation scenarios with the depletion they would cause for different crust compositions.  For primordial composition of the lunar crust, we take estimates of sodium and potassium abundances from values observed in Ferroan Anorthosite (FAN) samples\cite{2014RSPTA.37230241R, 2014E&PSL.388..318G, doi:10.1111/j.1945-5100.2010.01067.x, doi:10.1111/j.1945-5100.2009.tb01223.x}, which we use to represent the abundances expected from the primary lunar crust\cite{doi:10.2138/am-2015-4817}.  Values for Sodium and Potassium are 0.26\% (taken from a 0.35\% Na$_{2}$O abundance by weight) and 0.026\% (similarly calculated from the weight percentage of K$_{2}$O).  Total depletion of the regolith is then treated as an incompatible space weather and solar rotation state scenario given the non-negligible abundances of both elements observed in the regolith\cite{Mckay_7the, 2013NatSR...3E1611Z}.  Total depletion in figures \ref{fig:cme_frequency_timeNa} - \ref{fig:cme_frequency_timeK90} are given as a percentage of the total fraction of the abundance of that particular element in the FANs sample. We include a line for the abundance in FANs (corresponding to 100$\%$ since we use FANs for normalization) and also include a line for abundances in the present day Earth continental crust\cite{2003TrGeo...3....1R} (an unrealistic very high bound used for comparison). 

The lower values of sodium and potassium in FANs reflect the observed depletion across more volatile elements on the Moon versus the Earth.  Differences in bulk properties between the two bodies are attributed to differences in formation and evolution between the two bodies, and while post-formation depletion of moderate volatiles may play some role in this, some combination of preferential accretion of refractories and outgassing of volatiles during formation stages still appears to be the best explanation for the bulk of depletion relative to values on Earth.  However, in order to remain agnostic about the timing of such loss, we compare potential depletion to a hypothetical and somewhat unrealistic upper bound, the present Earth continental crust abundance values\cite{2003TrGeo...3....1R}.  This value is merely meant to be informative (and for example, is greater than moderate volatile abundances in the Akali Suite) to compare total loss.  The total reservoir of sodium and potassium in the reservoir of the regolith is given in table \ref{table:volres} and is used to produce the depletion figures.  In that table, the large regolith values take the larger regolith depths, the small regolith values use the smaller depths and the PEC large regolith reflects a present Earth Continental Crust abundance.

\section{Results} \label{sec:Results}

\begin{table}[h]
\caption{Moderate Volatile Loss Estimates for Solar Rotation Histories}
\begin{adjustbox}{width=0.95\textwidth}
\begin{tabular}{|c|c|c|c|c|}
\hline
(Values in $1^{14}$ kg)  & Total ($10^{32}$ erg CMEs) & Total ($10^{32}$+$10^{33}$ erg CMEs) & \textless{}1 Gyr ($10^{32}$ erg CMEs) & \textgreater{}1 Gyr ($10^{32}$ erg CMEs) \\
\hline
Slow Rotator (Na)       & 8.1                             & 9.7                                     & 6.9                               & 1.2                                  \\
\hline
Medium Rotator (Na)     & 11.9                            & 14.3                                    & 10.7                              & 1.3                                  \\
\hline
Fast Rotator (Na)       & 1300                            & 1600                                    & 1300                              & $\sim$1                              \\
\hline
\hline
Slow Rotator (K/50\%)   & 2.4                             & 2.8                                     & 2.0                               & 0.4                                  \\
\hline
Medium Rotator (K/50\%) & 3.5                             & 7.0                                     & 3.1                               & 0.4                                  \\
\hline
Fast Rotator (K/50\%)   & 380                             & 770                                     & 380                               & $\sim$0.3                            \\
\hline
\hline
Slow Rotator (K/95\%)   & 4.5                             & 5.4                                     & 3.8                               & 0.7                                  \\
\hline
Medium Rotator (K/95\%) & 6.6                             & 13                                      & 5.9                               & 0.7                                  \\
\hline
Fast Rotator (K/95\%)   & 730                             & 1500                                    & 730                               & $\sim$0.6                         \\
\hline
\end{tabular}
\end{adjustbox}
\label{table:lossest}
\end{table}

In the cases shown in figures \ref{fig:cme_frequency_timeNa}-\ref{fig:cme_frequency_timeK90}, we assess total depletion for the slow, medium and fast rotator cases (see table \ref{table:lossest} in appendix).  In each case, a Fast Rotator Sun would have depleted all of the given elements from a FAN composition regolith by orders of magnitude and even would have completely depleted an enriched Earth continental crust abundance regolith.   Given observed regolith abundances of sodium and potassium\cite{Mckay_7the}, this is clearly not the case.  These joint constraints thus strongly suggest that the Sun was not a fast rotator.  The other consistent patterns from the figures are that in each case it is evident that the majority of depletion occurred prior to 3.5 Gyr ago - before emplacement of the largest volume of mare basalts.  Only potassium depletion may have exceeded 10$\%$ after that period. Depletion increases with solar rotation speed (see figure \ref{fig:cartoonSuns} for a cartoon description) and after 3.5 Gyr total depletion between cases remains about the same (as Solar rotation rate converges - see figure \ref{fig:cme_frequency_time}). 

While sodium depletion is less than the total FAN abundance in both the slow ($\sim10-30\%$) and medium rotator case ($\sim15-40\%$), both of the graphs depicting potassium loss suggest that a medium rotator Sun would have depleted the total FAN content of potassium.  This suggests there is some evidence that the Sun was more likely a slower rotator.  This is more complicated to assume given that additional emplacement likely occurred during volcanic episodes and then was mixed into the regolith.  However, given the relatively small fractional volume of even high aklali suite and KREEP surface composition relative to FAN composition\cite{Wang2015}, this was unlikely to enhance total abundance past the predicted medium rotator depletion values.  Additionally, given the mostly conservative assumptions made in the model with respect to loss, these results likely understate the expected depletion in all the rotator cases. Finally, a significant fraction of the total FAN abundance for both elements is depleted by the solar activity.  Even in the slow rotator case for sodium, a total of about 10-30$\%$ is depleted due to solar activity.  Given abundances of sodium and potassium in FANs, the depletion values are potentially measurable (though regolith mixing and other processes contributing to variations may make interpretation complicated).  Potassium does seem to offer a better chance at observing 'recent' (post 3.5 Gyr) depletion from all these cases, with high escape fraction potassium cases being depleted by 10s of percent.

\begin{figure}[h!]
    \centering
    \begin{minipage}{0.50\textwidth}
        \centering
        \includegraphics[width=\textwidth]{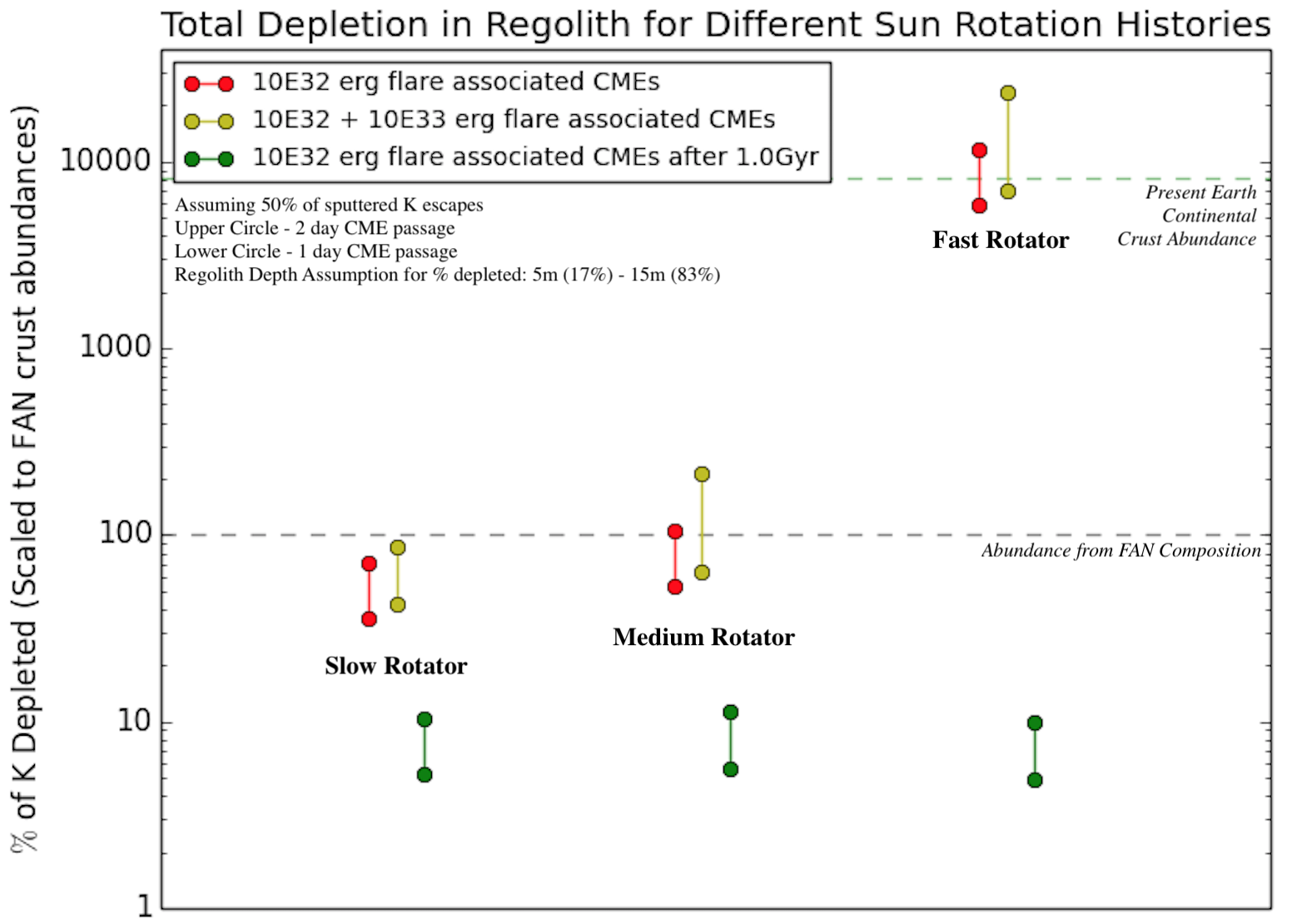} 
        \label{fig:cme_frequency_timeK50}
    \end{minipage}\hfill
    \begin{minipage}{0.50\textwidth}
        \centering
        \includegraphics[width=\textwidth]{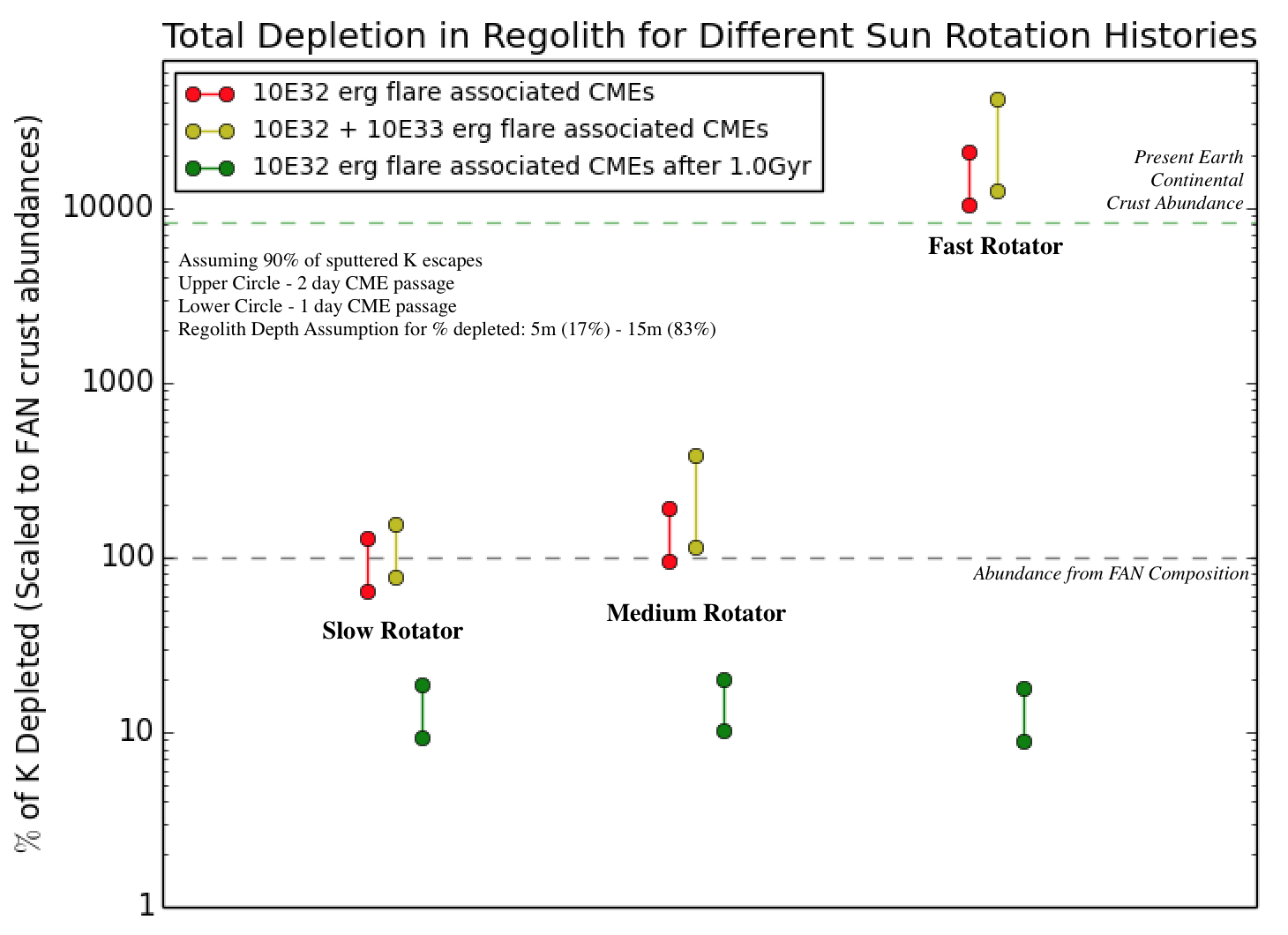} 
        \label{fig:cme_frequency_timeK90}
    \end{minipage}
    \caption{\textit{Depletion of potassium from the lunar regolith for different solar rotation cases. Different color values and crust assumptions/composition are labeled similarily to the sodium figure.  The left/right images are lower/upper bounds which assumes 50\%/90\% of sputtered potassium escapes.}}
    \label{fig:cme_frequency_timeK90}
\end{figure}

\section{Additional Relevant Factors: Regolith Exposure Time, Impacts, Volcanism and Magnetism} \label{sec:Discussion}

\textbf{Regolith Exposure Time}: Given the high rate of depletion the surface may have experienced in the 1st Gyr due to the high frequency of incident CMEs, it is important to examine whether the regolith was completely depleted of the relevant elements on short enough time scales to prevent further loss. In order to ensure depletion did not outpace the total content of the accessible reservoir during the most active periods of sputtering for any of the rotator cases, we examined the total volume of material exposed versus depleted.  Using an updated model for regolith overturn\cite{COSTELLO2018327}, which importantly considers the effect of secondaries in order to estimate the turnover time for a depth, we find that a 1 cm depth would be incredibly well mixed in 10,000 years ($>$100 times to 99$\%$ probability).  Over the most active 10,000 year period for our solar rotation cases, the total reservoir of sodium in that layer would be about $\sim$20-30 times greater than the total depleted by CMEs for the slow and medium rotator cases and $\sim$2 times greater even in the fast rotator case. This still understates how unlikely depletion of the sputtering depth reservoir is, as $>$100 overturns is very well mixed and depths down to $\sim$3-4 cm are overturned at least once over the same period.  Additionally, the calculations use impactor flux rates from the past billion years to inform the overturn rate, which are known to be far smaller than the more active impact environment of the early Moon that is relevant in this case.

\textbf{Impacts}: In addition to churning the regolith, impacts may also have influenced composition changes.  In the case of the maximum enrichment impacts could have caused (thus allowing greater depletion), using a 2$\%$ by weight contribution of meteoritic material to the regolith\cite{doi:10.2138/rmg.2006.60.1, 1987JGR....92..447K} and taking abundances from the most volatile rich chondrites\cite{10.2307/37983}, the enrichment does not change any of the previously mentioned conclusions. Conversely for large impacts we find that except for the heaviest, fastest, and most iron dominated impactors, the cumulative impacts influenced abundance changes less significantly by orders of magnitude when scaled to impact models\cite{doi:10.1146/annurev-earth-063016-020131} and the largest known basins or total meteoritic content of the regolith.  Further, additional depletion by these hypothetical events would only further restrict the likelihood of a fast/medium rotator Sun because those cases would more easily deplete the lesser abundances.  
The method we used for the impact vaporization is also given in Killen et al. (2012) with the modification such that for individual impacts a given impact velocity was specified. Therefore we did not integrate over a velocity distribution nor did we integrate over a size distribution since the impactor radius and mass were specified.  With the exception of a massive mantle breaching and emplacing impact such as the one that may have produced the South-Pole Aitken (SPA) Basin, all the impacts we tested would only vaporize moderate volatiles that would further drive depletion, given the energetics of the impactor.  In the case of a SPA Basin associated event, even if several times the total amount of sodium and potassium in the regolith were added to the crust, the proposed early time of the impact would still enable a fast rotator Sun to deplete the emplaced moderate volatiles by several factors.   

\begin{figure}[t]
\begin{center}
\includegraphics[scale=0.25]{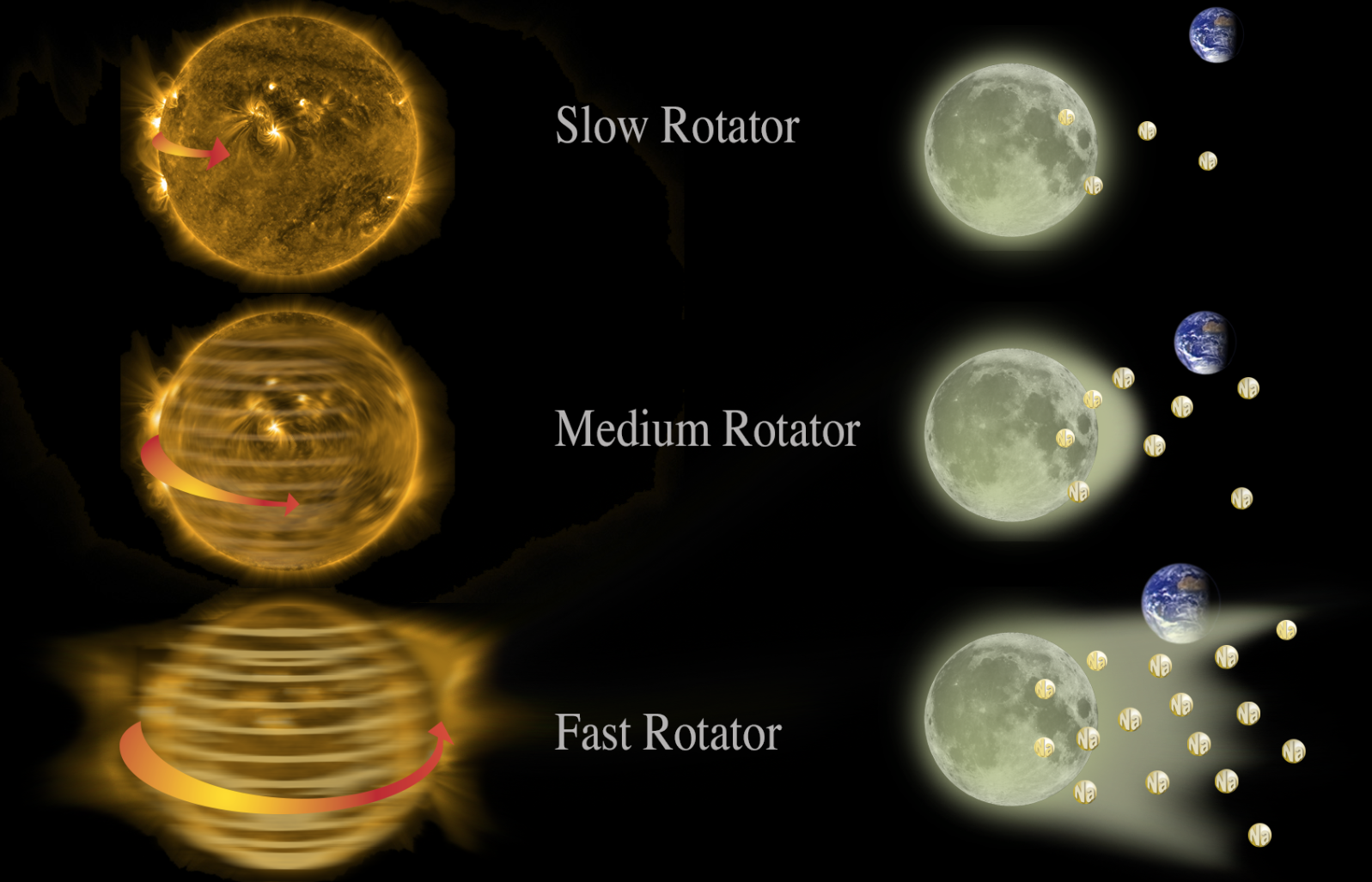}
\caption{\textit{A cartoon image depicting the change in moderate volatile loss from the surface of the Moon as a function of solar rotation speed. (NASA GSFC/Jay Friedlander)}}
\label{fig:cartoonSuns}
\end{center}
\end{figure}

\textbf{Volcanism}: Volcanism is also unlikely to impact the main conclusions of our study.  The bulk of mare emplacement occurred after the period of both the highest solar activity and impacts on the Moon\cite{NEEDHAM2017175}.  While the total mass of sodium emplaced may have been comparable to the total that would have been lost due to a fast rotator Sun, this would have to assume that the sodium was well mixed laterally over the Moon's surface and that the abundance was fairly high ($>.50\%$).  The same is also true of potassium.  However, there are two reasons that this was highly unlikely to be the case.  First, the vast majority ($>95\%$) of the basalt was emplaced at depths\cite{NEEDHAM2017175} that far exceed the total depth to which the regolith has been churned in the mare regions\cite{Mckay_7the}.  Without being excavated to greater depths, there is an insufficient mass of sodium and potassium in the relatively shallow reservoirs to explain abundances observed in non-mare regions.  Secondly, maps of potassium abundance\cite{2013NatSR...3E1611Z} on the Moon do show the expected increased abundance in mare regions but also show that the abundance sharply (with fairly stable values in highland regions) falls off away from this region, suggesting limited lateral mixing of material from the region with the rest of the surface.   Given the fall off in impact rate at this relatively late stage and the relatively shallow depths to which the mare regolith appears to be churned, it is unsurprising that geographically isolated chemically distinct regions have remained fairly well separated.

\textbf{Lunar and Terrestrial Magnetic Field Effects}: 
Finally, we elaborate on the potential effect of magnetic fields - an important topic considering the increasing amount of evidence suggesting the Moon had a moderately strong magnetic field early in its' history.  The Moon may have possessed a magnetic field of strength perhaps up to several tens of $\mu$T until about 3.5 Gyr ago and then an order of magnitude weaker for an additional period of time\cite{Weiss1246753, Garrick-Bethell356}.  However, the frequency of the much stronger CMEs, and increased ram pressure from what were essentially a continuous stream of powerful CMEs with increased particle densities and velocities (see table \ref{table:solarprop}) during peak activity (with $>$1 day passage times and $\sim$ 1/day frequencies) would have enabled fairly common access to the surface. Such incident space weather would have meant required relaxation times for compressed magnetic fields may not have been sufficient to prevent shifting of open-closed magnetic field line boundaries\cite{2016NatGe...9..452A}. These stronger, successive storms would thus allow access to a significant portion of the surface\cite{2014NatCo...5E3481L}. Consequently, a strong magnetic field is unlikely to have prevented depletion by a fast rotator.  

In the case of a medium rotator, a strong magnetic field may have had a more significant shielding effect given the stochastic nature of the CME impacts, but even in that case, reduction of the total depletion of potassium predicted by the model may not be enough to allow for such a scenario.  However, this interaction is a fairly important factor that should be studied given its potential to influence escape and transport of material on the early Moon.  Additionally, the relatively commonplace impact of several consecutive powerful CMEs during periods of lava emplacement may explain high magnetization values observed in lunar samples that are difficult to explain by most dynamo models.  This may have been a higher energy counterpart to a potential effect noted\cite{2018LPI....49.2439G} for a stronger earlier solar wind. 

The Earth's magnetosphere is also unlikely to have significantly shielded the Moon for the findings relevant to this study.  Given the quick migration of the Moon away from the Earth post-formation (to  $\sim80\%$ of its current separation by 100 Myr\cite{1994AJ....108.1943T}), even a generous 40-50$\%$ reduction in the amount of time the Moon would have been exposed to CMEs by the magnetosphere is insufficient to explain lack of depletion by a fast rotator and likely even a medium rotator Sun given the calculated results and observed abundances.  This reduction is an overestimate given that the Moon currently only spends about 20\% of it's orbital period in the Earth's magnetotail and given that the earlier orbital distance of $\sim48 R_{e}$ far exceeds the $\sim15 R_{e}$ extent of the magnetosphere along the Earth's day-night terminator (with an even smaller distance on the dayside).

\section{Conclusion and Potential Observational Signatures} \label{sec:Conclusion}

After considering potential complications, two underlying takeaways from the results remain.  First, that regardless of the initial rotation state of the Sun, the increased CME activity produced in the $\sim$1 Gyr after Moon formation led to a significant proportion of depletion of sodium and potassium from the surface given an assumption of FAN composition. Secondly, the abundances of sodium and potassium measured in regolith soil strongly suggest the Sun was not a fast rotator and in the case of potassium provides evidence that the Sun was likely a slow rotator.  This constrains the activity of the Sun not just in terms of CME frequency, but also in past solar wind and radiative properties. 

This history of solar activity is critical to constrain given that it would strongly influence the evolution of all the bodies in the inner solar system.   Indeed, given the methodology we used with respect to CME morphology assumptions, the CME incidence rate in figure \ref{fig:cme_frequency_time} would be the same rate for Mercury, Venus and Mars.  The main difference for these bodies would be CME properties including velocity and particle density along with potential interaction with evolving planetary magnetic fields.  In the cases of Venus and Mars, the implications are that this early period of activity may have significantly influenced those planets' atmospheric loss and chemistry in ways that deserve detailed study.  Mercury, however, may have had space weather interaction more analogous to that experienced by the Moon in the past if planetary conditions were similar to present day Mercury conditions - specifically, no atmosphere and a relatively weak global magnetosphere.  The processes that would have operated at the Moon would then potentially have also similarly affected Mercury in a heightened manner given its' proximity to the Sun.  However, at present there are no confirmed samples from Mercury which can be examined in a manner that is analogous to existing lunar samples.  Additionally, remote sensing observations of Mercury have indicated surprising abundances of moderate volatiles and water on the surface and poles (particularly in shadowed craters) of the planet that are somewhat puzzling given the thermal and space weather environment of the planet \cite{2014Icar..228...86P, 2017arXiv171202187N, 2013JGRE..118...26C}. However, leading hypotheses for these greater than expected abundances are some type of relatively recent exogenous source or volcanism \cite{1999Icar..137..197M, 2018Icar..302..191J} - meaning that extracting past space weather effects solely from remote observations may be difficult due to degeneracies from these potentially overlaid events/mechanisms.

The difficulty in assessing the specific nature of this past space weather environment using evidence from other solar system bodies makes potential evidence preserved in the lunar crust particularly valuable. What is especially notable is that there may be multiple independent observational signatures of the early space weather environment accessible on the Moon's surface.  The period of greatest depletion overlapped with the period of greatest impacts, and given that depletion monotonically declined after formation of the Moon's crust, such changes may be recorded with depth in the regolith as churn depths changed over time.  Since depletion should have been at least a quasi-global feature, such a stratigraphic signal may be preserved either in shielded samples (such as paleoregolith samples) or in a population of samples from geographically and temporally diverse regions.  Samples with different exposure ages that reflect the cumulative impact of different periods of space weather activity may provide chemical signatures with respect to abundances and fractionation that can help put tighter constraints on space weather and solar activity over time.  In addition to sputtering driven depletion, solar activity would also uniquely concurrently implant elements (for example, Argon) in the lunar crust that are most likely to have been sourced from solar activity. Thus potential lines of evidence may be different levels of depletion as a function of exposure (as explored in this study), implantation of uniquely exogenous ratios of specific elements such as Argon and Neon and potential evidence of fractionation in elements such as Nitrogen\cite{1975LPSC....6.2131B, 1975Sci...188..162K}. Thus cross cutting, diverse lines of chemical evidence can be used as a potential marker to constrain past solar activity in stratigraphy. 

Additionally, while most sputtered moderate volatiles would escape during CME passage, some proportion would return to the surface and would preferentially stick to colder surfaces towards the poles and 'permanently shadowed regions' (PSRs).  Given the increased loss early on, PSRs that existed prior to and after potential reorientation episodes of the Moon may exhibit higher abundances of these moderate volatiles (and may track reorientation episodes by capturing volatility influenced abundance gradients). In these cases, PSRs that remained largely shadowed over time may contain a vertical profile of deposited material (though again subject to mixing by regolith churn) that reflects volatile transport at least partially influenced by space weather activity.  This may also be true of ancient lava tubes which similarly would protect volatiles once they become trapped.  For PSRs that eventually became exposed to sunlight due to obliquity variations or reorientation, the total abundance and preferential loss of specific elements based upon their volatility may provide not only a proxy of the nature of reorientation events but also of deposition prior to and between these events.  This should be considered for future sample analysis and gamma ray spectrometer missions, and conveniently may be accessible in portions of the Moon where there is already high interest in landing.  Finally, along with a relatively high frequency of very strong CMEs, an early period of increased space weather activity would also have included more frequent very energetic CME events (at least an order of magnitude more energetic than the most energetic CMEs examined here).  Those CMEs may have induced spallation (either directly or through injected Solar Energetic Particle (SEP) events) that would have been recorded in 'fission tracks' in samples.  Indeed, such evidence has been used to examine potential changes in solar particle flux in relatively recent lunar history \cite{1980asfr.symp..201Z, 1979LPICo.390...17C}, but has been less well examined after the Apollo era.  Such particle track analysis may be an additional invaluable means of tracing solar activity history.  Given all the different processes which may have left signatures recording critical ancient and changing space weather, a more detailed study looking at the specific sample signatures, including potential changing inputs, is warranted.  The value of samples from different regions of the Moon is thus even more obvious, given that the history of the Sun is buried in the lunar crust.

\newpage
\appendix

\section{Additional Data Tables}

\begin{table}[h]
\centering
\caption{Physical Sputtering Yields}
\begin{tabular}{|c|c|c|c|c|}
\hline
(Loss in $atoms$ $cm^{-2}$ $s^{-1}$) & Sodium & Potassium & Magnesium & Calcium \\
\hline
Low Energy CME (C)                                           & 1.3E5  & 4.2E4     & 1.2E6     & 1.4E6   \\
\hline
Medium Energy CME (M)                                        & 9.5E5  & 3.1E5     & 8.7E6     & 1.0E7   \\
\hline
High Energy CME (X)                                          & 9.5E6  & 3.1E6     & 8.7E7     & 1.0E8   \\
\hline
Superflare associated CME (X10)                              & 1.9E8  & 6.2E7     & 1.8E9     & 2.1E9  \\
\hline
\end{tabular}
\label{table:sputterval}
\end{table}

\begin{table}[h]
\centering
\caption{Regolith Moderate Volatile Reservoir}
\begin{tabular}{|c|c|c|c|}
\hline
(Values in $1^{14}$ kg) & Large Regolith (FAN) & Small Regolith (FAN) & Large Regolith (PEC) \\
\hline
Sodium              & 33                   & 23                   & 296                  \\
\hline
Potassium           & 3.3                  & 2.3                  & 270    \\
\hline
\end{tabular}
\label{table:volres}
\end{table}

\textbf{Acknowledgements}: V.S.A. was supported by NASA grant \#80NSSC17K0463.  The authors would like to thank the anonymous reviewer whose suggestions helped improve the quality of the paper.

\bibliography{sample}




\end{document}